\documentstyle[12pt,graphics]{article}
\input epsf

\begin{document}
\baselineskip 24pt

\newcommand{\be}{\begin{equation}}
\newcommand{\ee}{\end{equation}}

\newcommand{\bea}{\begin{eqnarray}}
\newcommand{\eea}{\end{eqnarray}}

\newcommand{\ol}{\overline}

\newcommand{\sheptitle}
{ Implications of $B_s\to \mu^{+}\mu^{-}$ in 
  $SO(10)$-like Models}

\newcommand{\shepauthor}
{T. Bla\v{z}ek$^*$, S. F. King and J. K. Parry}

\newcommand{\shepaddress}
{Department of Physics and Astronomy, University of Southampton \\
        Southampton, SO17 1BJ, U.K}

\newcommand{\shepabstract}
{We examine the potentially very promising signal 
$B_s\to \mu^{+}\mu^{-}$ in supersymmetry 
with large $\tan\beta$ in a {\em top-down} approach starting
from the best fits of an $SO(10)$-like model studied recently.
Our results go beyond minimal
flavour violation investigated in previous works.
We show that the absolute best fits provide a signal for 
$B_s\to \mu^{+}\mu^{-}$ at the borderline of the
present limits and hence the ongoing search at the Tevatron will 
start having an impact on the global analysis of this class of 
SUSY models. We discuss the implications of a measurement of
$B_s\to \mu^{+}\mu^{-}$ for restricting the parameter space
of gauginos and sfermion masses, and of signals in other channels
$B_{d,s}\to \ell^{+}\ell^{-}$. We also discuss correlations
of $B_s\to \mu^{+}\mu^{-}$ with the CP-odd Higgs mass,
$\sin (\beta -\alpha)$ and $b\rightarrow s \gamma$ in $SO(10)$-like models.}

\begin{titlepage}
\begin{flushright}
hep-ph/0308068 \\
SHEP-03/26
\end{flushright}
\begin{center}
{\large{\bf \sheptitle}}
\\ \shepauthor \\ \mbox{} \\ {\it \shepaddress} \\ 
{\bf Abstract} \bigskip \end{center} \setcounter{page}{0}
\shepabstract
\begin{flushleft}
\today
\end{flushleft}

\vskip 0.1in
\noindent
$^*${\footnotesize On leave of absence from 
the Dept. of Theoretical Physics, Comenius Univ., Bratislava, Slovakia}

\end{titlepage}

\newpage

\section{Introduction}
Ideas of unification and the origin of flavour have been 
under investigation for a long time and many different 
models have been proposed in the last twenty years.  Yet  
in the diversity of different approaches a class of
unification models can be recognised which is remarkably
simple at the unification scale. We call this class
 SO(10)-like unification models. In these models the 
effective theory at the unification scale assumes that 
the Standard Model (SM) gauge couplings unify to a per cent 
level, third family yukawa couplings are all of
order unity and the remaining flavour structure originates 
in a small set of higher-dimensional superpotential 
operators keeping the supersymmetry (SUSY)
breaking sector of a model flavour blind. 
We note that actual models which fall into this category 
often assume lower symmetry than SO(10), e.g. models 
based on the Pati-Salam gauge group or the MSSM gauge
group generated by a string theory in higher dimensions 
are often found in this class of models. 

It has been recently pointed out that 
if the Minimal Supersymmetric Standard Model (MSSM) is the 
effective theory describing nature above the scale $100\,$GeV
and $\;\tan\beta\equiv\langle H_u^0\rangle\,/\,
                      \langle H_d^0\rangle
                \equiv v_u/v_d\:$   
is large, 
a pure leptonic $B_s\to\mu^+\mu^-$ decay has a very strong case to
emerge among the first indirect signals of supersymmetry 
(SUSY) \cite{Bs.to.mumu.general}. This  is because the decay signal 
should be very clear  at the Tevatron or LHC and 
also because the SM branching fraction is suppressed 
down to $10^{-9}$ while the rate can be enhanced when considering
SUSY extensions. In particular, this occurs
due to large couplings of the down-type quarks and charged leptons
to the MSSM higgs states if $\tan\beta$ is large. 
Thus it is important to analyse this decay in a full
SUSY theory and not just
in terms of the minimal flavour violation that assumes that
$V_{cb}$ is the only source of the $32$ transition, as has been done
in the past.
In the full context of complete unification models
it means that the $32$ flavour structure 
is restricted by the fermion mass ratios $m_\mu/m_\tau$,
$m_s/m_b$ and $m_c/m_t$, small value of $V_{cb}$,
large $U_{\mu 3}$, $b\to s\gamma$ branching ratio 
and possibly other low energy observables and constraints. 
Although these constraints do not determine the $32$ sector 
uniquely they do provide for a realistic prediction of
observables like the $B_s\to\ell^+\ell^-$ decay rates.

In this letter we present the results of such a complete {\em top-down}
investigation based on the best fit predictions obtained in a recent 
global analysis of a complete SO(10)-like model \cite{Blazek:2003wz}. 
The best fits obtained in this work give a very good agreement 
with the observables related to 32 flavour sector 
and satisfy all laboratory experimental constraints on superpartner 
masses. Here they serve as our starting point since they 
provide us with all the MSSM couplings at the low-energy scale. 
Within this framework we study the implications of 
a possible measurement $B_s\to\mu^+\mu^-$.
In particular, we discuss the related processes
$B_s\to\tau^+\tau^-$,
$B_d\to\mu^+\mu^-$, 
$B_d\to\tau^+\tau^-$,
and show the correlations with $B_s\to\mu^+\mu^-$.
We discuss the implications of a measurement of
$B_s\to \mu^{+}\mu^{-}$ for restricting the parameter space
of gauginos and sfermion masses, and also discuss correlations
of $B_s\to \mu^{+}\mu^{-}$ with $b\rightarrow s \gamma$ and
the CP-odd Higgs mass.

After this introduction the letter continues in section 2 with a brief
theoretical section on the evaluation of 
$B_s\to\ell^+\ell^-$ decay rate in {\it top-down} approach.
In section 3 we give some brief discussion of the 
$SO(10)$-like model we study. Section 4 contains our numerical
results, and a discussion of the implications of a signal for  
$B_s\to\mu^+\mu^-$ mentioned above. Section 5 concludes the paper.

\section{$B_s\to \mu^{+}\mu^{-}$}

We emphasise that in a {\it top-down} approach 
the tree-level MSSM couplings are determined from high energy boundary
conditions, and do not have to be determined by an 
iterative proceedure as in bottom-up 
approaches. In particular, in terms of effective vertices $f$ and $g$,
which are matrices in flavour space,
after heavy sparticles are integrated out 
the lagrangian can be written down as
\be
  {\mathcal L}_{eff} 
                       =
                         -\ol{d}_{R}^{(0)}
                          \left[
                              Y_d^{(0){\mathrm Diag}\:\dagger} H_{d}^{0}
                            + f^{\dagger}                    H_{d}^{0}
                            + g^{\dagger}                    H_{u}^{0\,*}
                          \right]
                             d_{L}^{(0)} + h.c..
\label{eq:L.eff.vert}
\ee
At tree level down-type quarks $d_{L,R}$
only couple to down-type Higgs $H_d^0$ and $f=g=0$.
Yukawa couplings $Y_d^{(0)}$ can be read out as a straightforward 
prediction of a unified model. 
$Y_d^{(0)}$ and the mass matrix $m_{d}^{(0)} = Y_d^{(0)}v_d$ can then be
simultaneously diagonalised with eigenvectors $d_{R,\,L}^{(0)}$. 
At one-loop level $f$ and $g$ have to be computed and 
the mass terms relevant for this discussion become 
\footnote{
          Terms due to wavefunction renormalisation do not
          contribute to flavour changing.
         } 
\be
  {\mathcal L}_{mass} 
                    =  -\ol{d}_{R}^{(0)}
                      \left[
                          m_{d}^{(0){\mathrm Diag}\:\dagger}
                        + f^{\dagger} v_{d}
                        + g^{\dagger} v_{u}
                      \right]
                         d_{L}^{(0)} 
\label{eq:L.mass}
\ee
using the same basis. Clearly, if $v_u\gg v_d$  sizeable 
corrections to the mass eigenvalues \cite{large.dmb} and mixing 
matrices \cite{Blazek:1995nv} are generated. Furthermore the 3-point 
functions in (\ref{eq:L.eff.vert}) and mass matrix in (\ref{eq:L.mass})
cannot be simultaneously diagonalised \cite{Hamzaoui:1998nu}.
If we write eq.~(\ref{eq:L.eff.vert}) as
\be
         -\ol{d}_{R}^{(0)}
          \left[
               Y_d^{(0){\mathrm Diag}\:\dagger}
            +  f^{\dagger}
            +  g^{\dagger} \frac{v_u}{v_d}
          \right]
                  d_{L}^{(0)}H_{d}^{0}
         \:-\:
           \ol{d}_{R}^{(0)}
           \left[ 
                g^{\dagger}
                       \left(
                          H_u^{0*}-\frac{v_u}{v_d}H_d^0
                       \right)
           \right]
                  d_{L}^{(0)}
\label{eq:L.eff.vert.2}
\ee
then the first bracket of eq.~(\ref{eq:L.eff.vert.2}) is in a form 
which is similar to that of the mass matrix and therefore is 
diagonal when $d_{L,\,R}^{(0)}$ are rotated into corrected
mass eigenstates $d_{L,\,R}^{(1)} = V_d^{L,R\,(1)} d_{L,\,R}^{(0)}$.
This is not true for the last bracket which becomes a source 
of flavour changing,
\be
    {\mathcal L}_{FCNC} \;=\; 
                            -\ol{d}_{R\,i}^{(1)} 
                             \left[
                                 V_{d}^{R\,(1)} g^{\dagger}
                                 \left(
                                       H_u^{0*}-\frac{v_u}{v_d}H_d^0
                                 \right)
                                 V_{d}^{L\,(1)\,\dagger}
                             \right]_{ij}
                             d_{L\,j}^{(1)} + h.c.,
\label{L_FCNC}
\ee
It is now explicit that its origin comes from
the interaction $\ol{d}_{R}^0H^{0*}_u d_{L}^0$, not present at tree level.
Moreover, the flavour changing couplings get enhanced by an explicit
factor $\tan\beta$ on top of any $\tan\beta$ scaling present in $g$.
In the leading order in $\tan\beta$ the $g$ matrix can 
in fact be related in a simple way to the
finite non-logarithmic mass matrix corrections,
$g_{ij} = (\delta m_d^{finite})_{ij}/v_u$,
computed for the first time in \cite{Blazek:1995nv}.
Due to 
$H_u^0=v_u+(H^0s_\alpha + h^0c_\alpha + iA^0c_\beta + iG^0s_\beta)/\sqrt{2}$ 
and
$H_d^0=v_d+(H^0c_\alpha - h^0s_\alpha + iA^0s_\beta - iG^0c_\beta)/\sqrt{2}$ 
we can write
\be
    H_u^{0*}-\frac{v_u}{v_d}H_d^0 \:=\:
                         \frac{1}{\sqrt{2}}\:\frac{1}{c_\beta}
                         \left[
                                 H^0 s_{\alpha -\beta} 
                               + h^0 c_{\alpha - \beta} 
                               -i\,A^0
                         \right],
\ee
where $s_\alpha\,\equiv\sin\alpha$, $c_\alpha\,\equiv\cos\alpha$, {\em etc}.
We can thus identify effective vertices $\ol{b}_Rs_LH^0$, 
$\ol{b}_Rs_Lh^0$ and $\ol{b}_Rs_LA^0$ involving $b$ to $s$ transitions
mediated by neutral physical higgs states. We note that with large 
$\tan\beta$ the coupling to the pseudoscalar $A^{0}$ is always
large while the CP-even states,
$h^{0}$ and $H^{0}$, have couplings which depend on the CP-even 
higgs mixing angle $\alpha$. The Goldstone 
mode is cancelled in the equation above and thus the effective vertex
with the $Z$ boson is absent at this level. 

In the MSSM with large $\tan\beta$ the dominant contribution 
to $B_s\to\ell^+\ell^-$ comes from the penguin diagram where 
the dilepton pair is produced from a virtual Higgs state 
\cite{Bs.to.mumu.general}. 
After the SUSY partners are integrated out we are left
with the effective vertices determined above. Thus
in combination with the standard tree-level term 
${\mathcal L}_{\ell\ell H} \:=\: -y_\ell \ol{\ell_R}\ell_LH_d^0 
                                 + h.c.$
the dominant $\tan\beta$ enhanced contribution to the 
branching ratio turns out to be
\bea
  BR(B^0_s\to\mu^+\mu^-) 
                     &=& 
                       2.25\times 10^{-3}\; 
                       \left| 
                             \frac{\delta m_{d\:32}^\dagger}
                                  {m_bV_{ts}}
                       \right|^2
                       \;
                       \left[ \frac{V_{ts}}{0.04} \right]^2
                       \left[ \frac{y_{\mu}}{0.0353} \right]^2
                       \left[ \frac{M_{170}}{v_u} \right]^2
                       \left[ \frac{\tan\beta}{50} \right]^2
                       \;
\times
\nonumber\\ 
              &\times &         
                       \left[
                          \left(
                              \frac{c_\alpha s_{\alpha-\beta}}
                                   {\left( \frac{M_{H^0}}{M_{100}}\right)^2}
                              -
                              \frac{s_\alpha c_{\alpha-\beta}}
                                   {\left( \frac{M_{h^0}}{M_{100}}\right)^2}
                          \right)^2
                          +
                              \frac{s_\beta^2}
                                   {\left( \frac{M_{A^0}}{M_{100}}\right)^4}
                       \right], \label{br}
\eea
where matrix $\delta m_d^\dagger$ is in the $\{d_{L,R}^{(1)}\}$ basis,
and is defined by
\be
\delta m_d^\dagger = V_d^{R\,(1)}(f^\dagger v_d + g^\dagger v_u)
V_d^{L\,(1)\dagger},
\ee
$m_b$ is the $b$ quark mass at scale $M_Z$ in the effective 
$SU(3)_c\times U(1)_{em}$ theory, the constants are  
$M_{100}=100\,$GeV and $M_{170}=170\,$GeV and
the numerical value is obtained from
\be
     2.25\times 10^{-3} \:=\:
                             \frac{\tau_B f_B^2 M_B^5}{128\pi}\;\,
             \frac{0.04^2\, 0.0353^2\, 50^2}{M_{100}^4\, M_{170}^2}.
\ee
Modification for other $B^0_{d_i}\to\ell^+\ell^-$ decays is trivial.
We note that each of these branching fractions actually scales down 
as $\tan^6\beta$ for lower values of $\tan\beta$:
additional powers of $\tan\beta$ enter 
due to the explicit presence of lepton yukawa coupling $y_\ell^2$
and  mass matrix corrections $\delta m_d^{finite}/m_b$ (or, equivalently,
yukawa coupling $y_{d_i}$ in $g$). 

\section{An $SO(10)$-like Model}

Our results are based on the model analysed in \cite{Blazek:2003wz}.
The model was defined below the $SO(10)$ breaking scale,
where the gauge group was broken to its maximal Pati-Salam
subgroup, and the flavour structure of the model was 
determined by operators which respected the Pati-Salam symmetry.
Universal gaugino masses $M_{1/2}$ and sfermion masses $m_F$
were assumed, and we allowed for $D$-terms and non-universal Higgs masses.
Throughout this work the trilinear parameter was kept fixed at $A_{0}=0$.
More details concerning the model can be found in \cite{Blazek:2003wz},
however the Yukawa matrices which enter at the unification scale
are listed below for completeness:

\bea
Y_u(M_{GUT}) & = & \left(\matrix{
                      \sqrt{2}\:a_{11}^{\prime\prime}\lambda^8 & 
                        \sqrt{2}\:a_{12}^{\prime}\lambda^5 & 
{\displaystyle \frac{2}{\sqrt{5}}}\:a_{13}^{\prime}\lambda^3   \cr
%
                                                       0     & 
{\displaystyle \frac{8}{5\sqrt{5}}}\:a_{22}^{\prime}\lambda^4 &
                                                       0     \cr
                                                       0     & 
{\displaystyle \frac{8}{5}}\:a_{32}^{\prime}\lambda^4 &
                                                       r_ta_{33}  \cr}
                              \right)
\nonumber \\
\noalign{\bigskip}
Y_d(M_{GUT}) & = & \left(\matrix{
             {\displaystyle \frac{8}{5}}\:a_{11}\lambda^6 &
          -\sqrt{2}\:a_{12}^{\prime}\lambda^5 & 
{\displaystyle \frac{4}{\sqrt{5}}}\:a_{13}^{\prime}\lambda^3   \cr
%
{\displaystyle \frac{2}{\sqrt{5}}}\:a_{21}\lambda^5 &
{\displaystyle \sqrt{\frac{   2}{  5}}}\:a_{22}\lambda^3   +
 {\displaystyle \frac{16}{5\sqrt{5}}}\:a_{22}^{\prime}\lambda^4 &
 {\displaystyle \sqrt{\frac{   2}{  5}}}\:a_{23}^{\prime}\lambda^2           \cr
 {\displaystyle \frac{8}{5}}\:a_{31}\lambda^6 &
     \sqrt{2}\:a_{32}\lambda^3 &
                                                        r_ba_{33}   \cr}
                                  \right)
\nonumber \\
\noalign{\bigskip}
Y_e(M_{GUT}) & = & \left(\matrix{
  {\displaystyle \frac{6}{5}}\:a_{11}\lambda^6 &
                                                        0    & 
                                                        0    \cr
%
{\displaystyle \frac{4}{\sqrt{5}}}\:a_{21}\lambda^5 &
-3\,{\displaystyle \sqrt{\frac{  2}{  5}}}\:a_{22}\lambda^3   +
{\displaystyle \frac{12}{5\sqrt{5}}}\:a_{22}^{\prime}\lambda^4 &
-3\,{\displaystyle \sqrt{\frac{  2}{  5}}}\:a_{23}^{\prime}\lambda^2       \cr
{\displaystyle \frac{6}{5}}\:a_{31}\lambda^6 &
       \sqrt{2}\:a_{32}\lambda^3 &
                                                        a_{33}    \cr}
                                  \right)
\nonumber \\
\noalign{\bigskip}
Y_{\nu}(M_{GUT}) & = & \left(\matrix{
     \sqrt{2}\:a_{11}^{\prime\prime}\lambda^8 & 
                     2 \:a_{12}\lambda^4 & 
                                                       0     \cr
%
                                                       0     & 
{\displaystyle \frac{6}{5\sqrt{5}}}\:a_{22}^{\prime}\lambda^4 &
                                      2 \:a_{23}\lambda            \cr
                                                       0     & 
 {\displaystyle \frac{6}{5}}\:a_{32}^{\prime}\lambda^4 &
                                                       r_{\nu}a_{33}     \cr}
                              \right)
\nonumber
\eea
where $\lambda=0.22$ is the Wolfenstein parameter, and
$a$ and $r$ are order unity coefficients which are precisely
determined in the global fit to give excellent agreement
with the observed quark and lepton masses and mixing angles.
The numerical Clebsch factors are shown explicitly.
Yukawa unification is not exact, with $r_b$ for example dropping
down to $0.7$ for the best fits, although we keep 
$\tan \beta =50$ fixed in our analysis.

The essential features of the flavour theory clearly include a large
off-diagonal neutrino Yukawa coupling $Y^{\nu}_{23}\sim 1$,
to generate the large atmospheric mixing angle, however
in the quark sector the Yukawa matrices have small off-diagonal
entries, and are not required to be symmetric.
The flavour structure of this model is therefore typical of many 
$SO(10)$ models, and has no particularly unusual features,
although of course we cannot claim it is generic since each
$SO(10)$-like model will differ in the details of its flavour structure.

\section{Results}

We first summarise the experimental limits for the processes of interest:
\begin{eqnarray}
{\rm Br}(B_{s} \to \mu\mu) & < &  2.0 \times 10^{-6}\, [{\rm CDF}]\\
{\rm Br}(B_{d} \to \mu\mu) & < &  6.1 \times 10^{-7}\, [{\rm Babar}]
\end{eqnarray}
with no bounds yet established for the $\tau$ final state processes.
Looking to the future, the Tevatron will bring us further results for 
$B_{s}$ decays with the prospect of a CDF bound in the 
region of Br$(B_{s}\to \mu\mu)<10^{-7}$.  
By comparison the standard model predicts 
${\rm Br}(B_{s} \to \mu\mu)_{\rm SM}\sim 3.5\times 10^{-9}$
{\cite{Buras:1998ra}}.

To obtain predictions for such processes, we have performed
a {\it top-down} global analysis of the $SO(10)$-like model outlined
in the previous section. The analysis \cite{Blazek:2003wz}
yields two distinct best fits, which
we call Minimum A and Minimum B. The higgs spectrum in Minimum B 
is heavy, mostly above the TeV scale and will not be considered
in the discussion below. The Higgs spectrum of Minimum A was found 
to be more interesting for our present study with masses at the 
$100$ GeV scale. Hence it is 
the results from the unaltered fits of Minimum A which we present 
in this letter.

\begin{figure}[p]
\begin{center}
\scalebox{0.75}{\includegraphics*{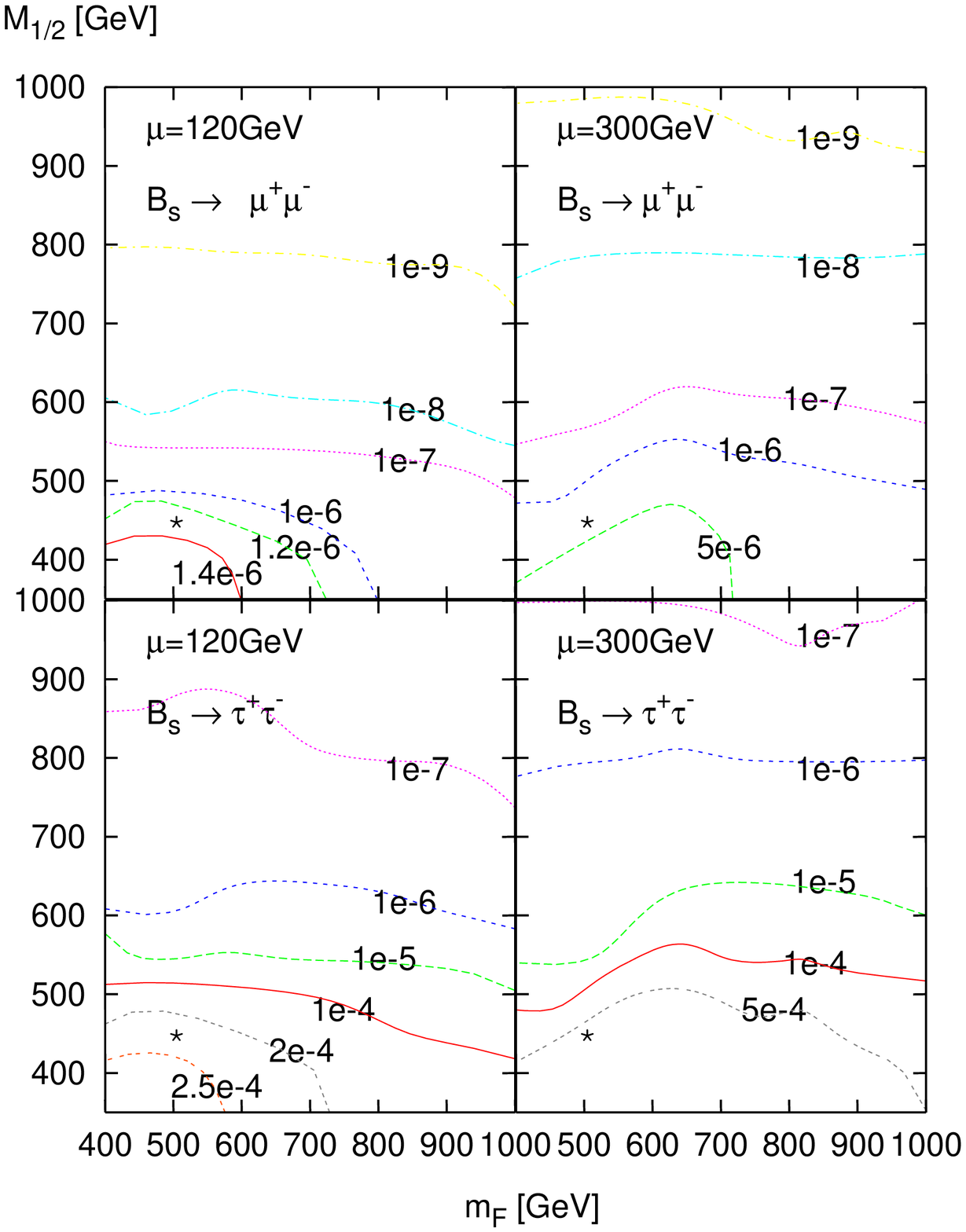}}
\begin{minipage}[t]{15cm}
\caption{\small{Contour plots for the branching ratios of the FCNC processes,
$B_{s}\to\mu^{+}\mu^{-}$ and $B_{s}\to\tau^{+}\tau^{-}$. Each branching ratio
is plotted with two different values of the $\mu$ parameter. The $\star$ marks
the best fit point. }}\label{Bs_plot}
\end{minipage}
\end{center}
\end{figure}

\begin{figure}[p]
\begin{center}
\scalebox{0.75}{\includegraphics*{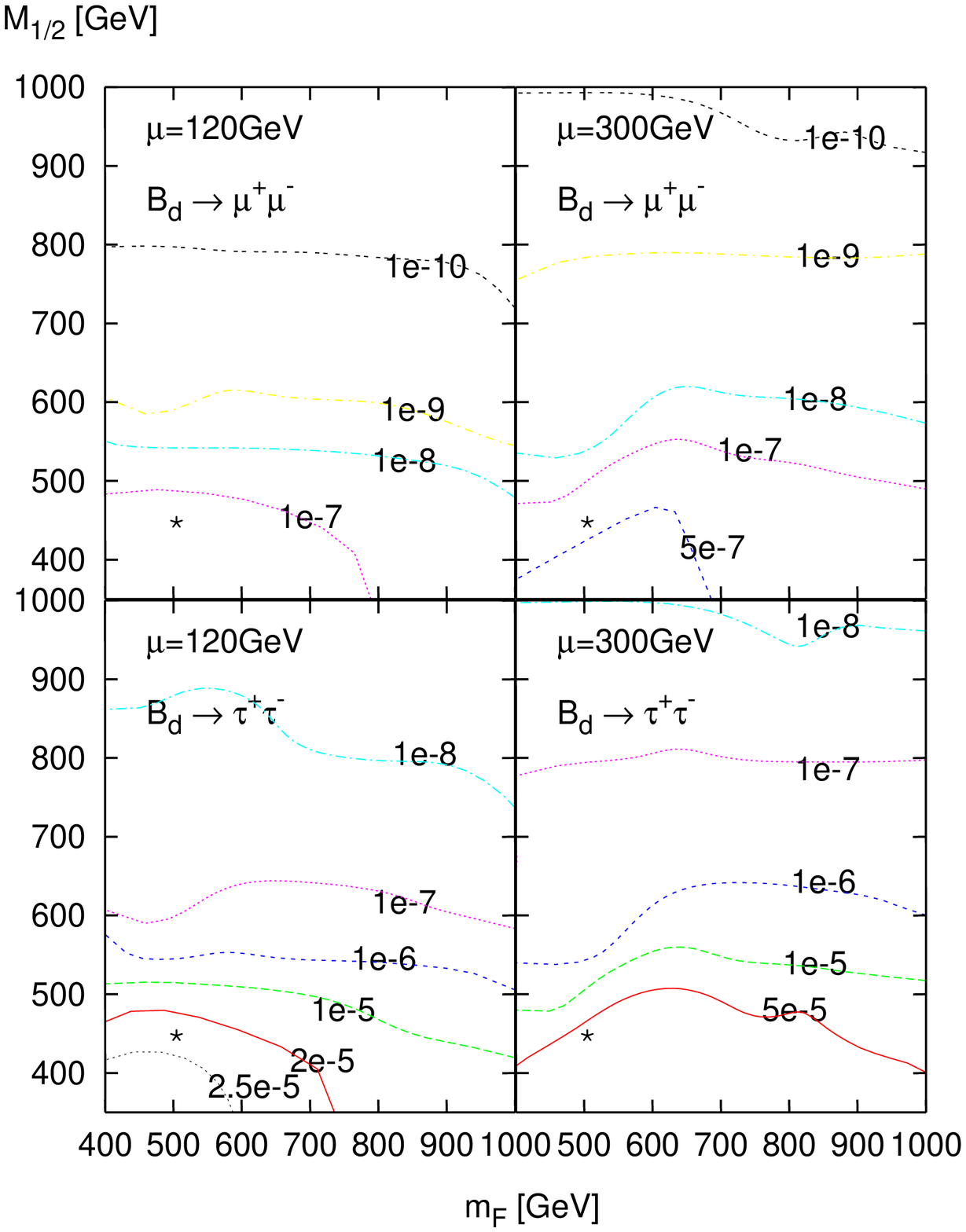}}
\begin{minipage}[t]{15cm}
\caption{\small{Contour plots for the branching ratios of the FCNC processes,
$B_{d}\to\mu^{+}\mu^{-}$ and $B_{d}\to\tau^{+}\tau^{-}$. Each branching ratio
is plotted with two different values of the $\mu$ parameter. The $\star$ marks
the best fit point.}}\label{Bd_plot}
\end{minipage}
\end{center}
\end{figure}

The numerical results for the processes 
$B_{s}\to \mu^{+}\mu^{-}$ and $B_{s}\to \tau^{+}\tau^{-}$ 
are displayed in fig.~{\ref{Bs_plot}}.
Similar results for $B_{d}\to \mu^{+}\mu^{-},\tau^{+}\tau^{-}$ 
are given in fig.~{\ref{Bd_plot}}. 
These results are presented as contour plots in the 
$m_{F}-M_{1/2}$ plane with a fixed value of $\mu=120$ GeV(left panels) 
and $\mu=300$ GeV(right panels). When comparing these contours with 
eq.~\ref{br} we 
find that a significant suppression is obtained from the ratio 
$\delta m_{d\,32}/m_{b}V_{ts}$. This comes purely from fitting the 
$b$ quark mass, $V_{cb}$ and $b \rightarrow s \gamma$.

The upper two panels of fig.~{\ref{Bs_plot}}
display contours of ${\rm Br}(B_{s} \to \mu\mu)$ with
$\mu=120$ and $300$ GeV, and show values quite close
to the current limits, and well above the standard model predictions.
The Higgs mediated contribution in the SUSY model clearly dominates 
over the standard model contribution
and for $\mu=300$ GeV, with low $M_{1/2}$, 
it can even exceed the present CDF limit. 
An improved limit of $10^{-7}$ would be very restricting 
and could probe Higgs
masses into the range, $m_{A^{0}}=150-300$~GeV. As for the process, 
$B_d \to \mu\mu$, fig.~{\ref{Bd_plot}} shows that the present bound 
is satisfied by both $\mu$ values over the entire displayed plane.

\begin{figure}[p]
\begin{center}
\scalebox{0.75}{\includegraphics*{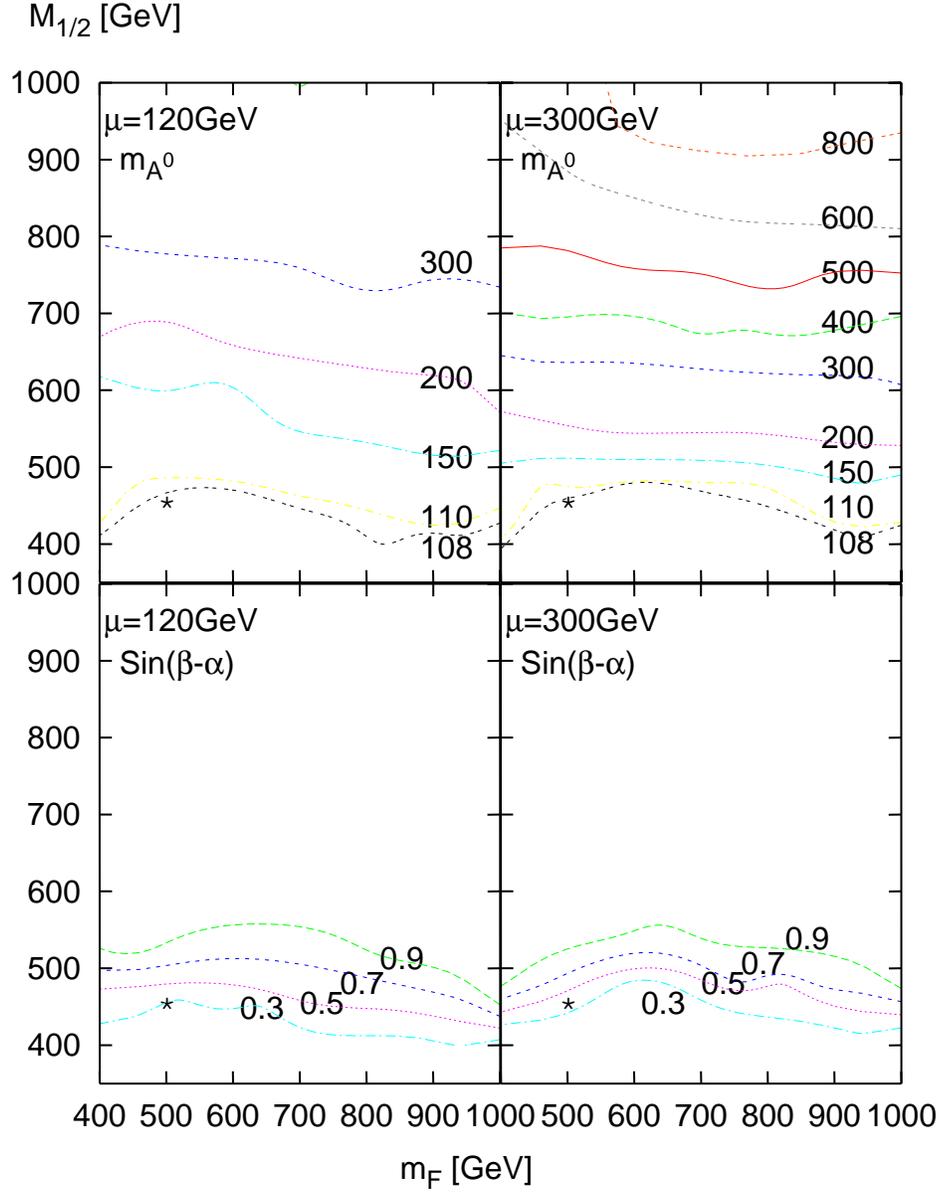}}
\begin{minipage}[t]{15cm}
\caption{\small{The upper two panels contain contours of the CP odd 
Pseudoscalar Higgs mass, plotted in the $m_{F}-M_{1/2}$ plane. 
The lower panels contain contours of, $\sin (\beta - \alpha )$, which determines
the strength of the Z-boson coupling to, $h^{0}$, the lighter CP even Higgs.
Again the plots are displayed at different values of $\mu$. The $\star$ 
marks the best fit point.}}\label{A0_Z_plot}
\end{minipage}
\end{center}
\end{figure}

Inspection of figs.~{\ref{Bs_plot}} and
{\ref{Bd_plot}}, reveals that the branching ratios for 
$B_{s,d}\to \mu^{+}\mu^{-},\tau^{+}\tau^{-}$
are sensitive to the universal gaugino mass $M_{1/2}$, but not to 
the universal sfermion mass $m_{F}$.
Inspecting the $m_{A^{0}}$ panels of fig.~\ref{A0_Z_plot} we see that it 
has a very similar $M_{1/2},\,m_{F}$ dependence. 
This is exactly as expected with a lighter mediating Higgs 
leading to larger branching ratios.

The branching ratio for $B_{s,\,d}\to \tau \tau$ is enhanced by
a factor of $(y_{\tau}/y_{\mu})^2 \sim 100$ compared to the muon final state
processes, as can be seen in the 
lower panels of fig.~{\ref{Bs_plot}} and {\ref{Bd_plot}}. This makes the tau 
final state processes very attractive for experimental discovery.
The difficulty comes with the required detector resolution to measure
tau decays. If this problem could be solved at future experiments then
these tau final state processes could become the primary signal for
indirect SUSY searches. 

Fig.~\ref{A0_Z_plot} contains corresponding contours of 
$m_{A^{0}}$ in the upper panels
and the quantity $\sin (\beta - \alpha)$, which controls the coupling of the
lightest CP-even Higgs scalar coupling to the Z,
in the lower panels.
The numerical predictions for the best fit point
at $M_{1/2}=450,\,m_{F}=500$ GeV 
(indicated by an asterisk in the figures)
are given in Table~{\ref{out_tbl}}.

\begin{table}[ht]
\begin{center}
\begin{tabular}{|c||c|c|}
\hline
             & $\mu=120$ GeV & $\mu=300$ GeV \\
\hline
\hline
$M_{1/2}$ [GeV]   &   $450$     &$450$     \\
$m_{F}$ [GeV]   &   $500$     &$500$     \\
\hline
$B_{s}\to \mu\mu $   &$  1.5 \times 10^{-6} $ & $  5.9\times 10^{-6}  $  \\

$B_{s}\to \tau\tau $   &$  2.6 \times 10^{-4} $ & $ 1 \times 10^{-3}  $  \\

$B_{d}\to \mu\mu $   &$   1.5 \times 10^{-7} $  & $  5.8 \times 10^{-7} $ \\

$B_{d}\to \tau\tau $   &$  2.7  \times 10^{-5} $ & $ 1 \times 10^{-4}$\\
\hline
$m_{A^{0}}$ [GeV]   &$  102           $ & $    102       $ \\

$\sin (\beta - \alpha )$   &$  0.22          $ &$    0.15     $   \\
\hline
\end{tabular}
\caption{Table of branching ratios for 
$B_{s,d}\to \mu^{+}\mu^{-},\tau^{+}\tau^{-}$,
CP-odd pseudoscalar mass $m_{A^{0}}$, and $\sin (\beta - \alpha )$
which governs the lightest CP-even scalar coupling to the $Z$,
for the best fit point.}\label{out_tbl}
\end{center}
\end{table}

\begin{figure}[p]
\begin{center}
\scalebox{0.75}{\includegraphics*{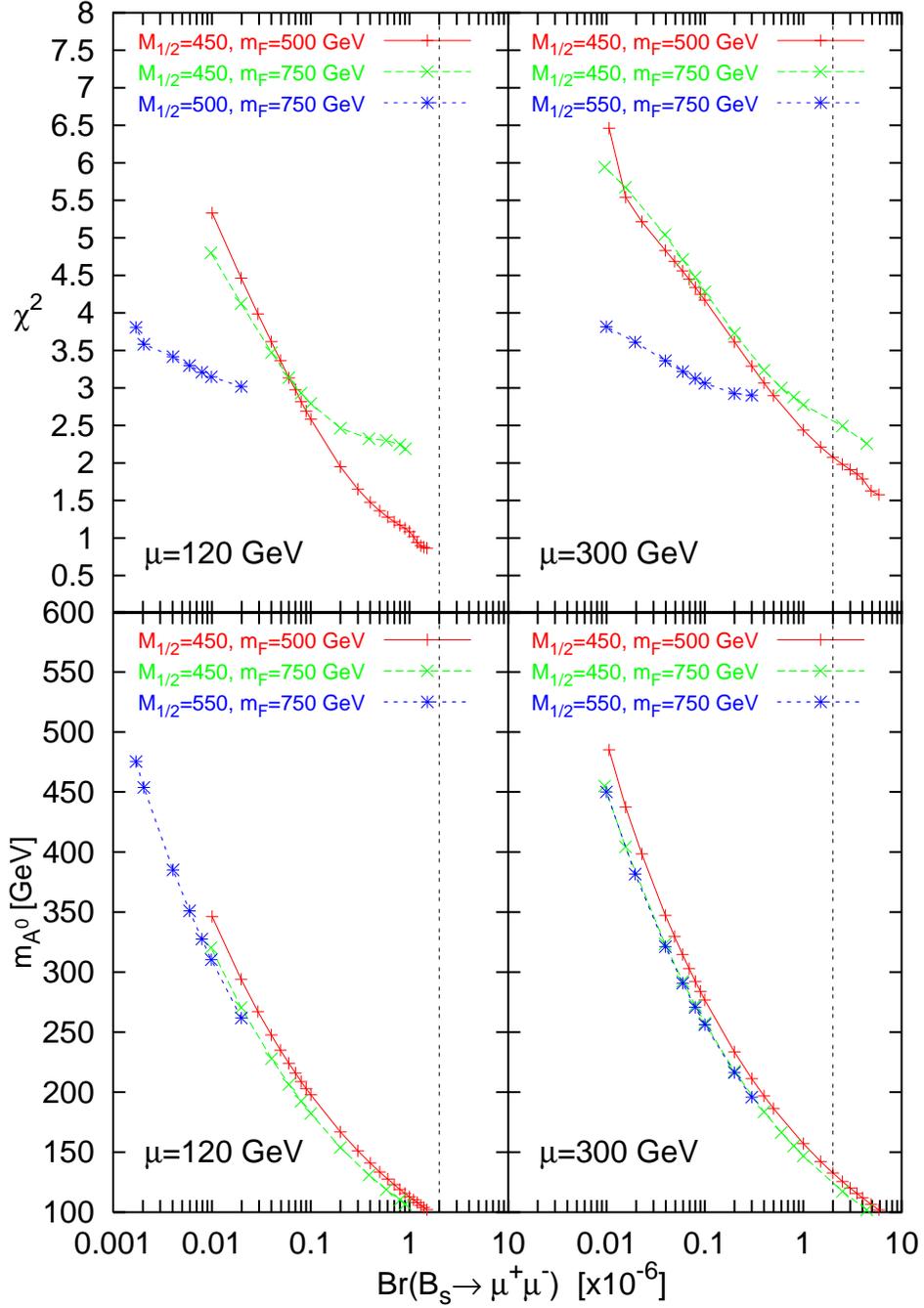}}
\begin{minipage}[t]{15cm}
\caption{\small{This figure shows the variation of
$\chi^2$ and the Pseudoscalar Higgs mass, $m_{A^{0}}$, as the branching ratio
 of $B_{s}\to \mu \mu$ varies from $10^{-5}$ down to $10^{-8}$. Each of the 
three curves are drawn with fixed values of $M_{1/2},\,m_{F}$. 
The vertical dashed line represents the present CDF
bound on $B_{s}\to\mu^{+}\mu^{-}$.}}\label{bsmm_var}
\end{minipage}
\end{center}
\end{figure}

\begin{figure}[p]
\begin{center}
\scalebox{0.75}{\includegraphics*{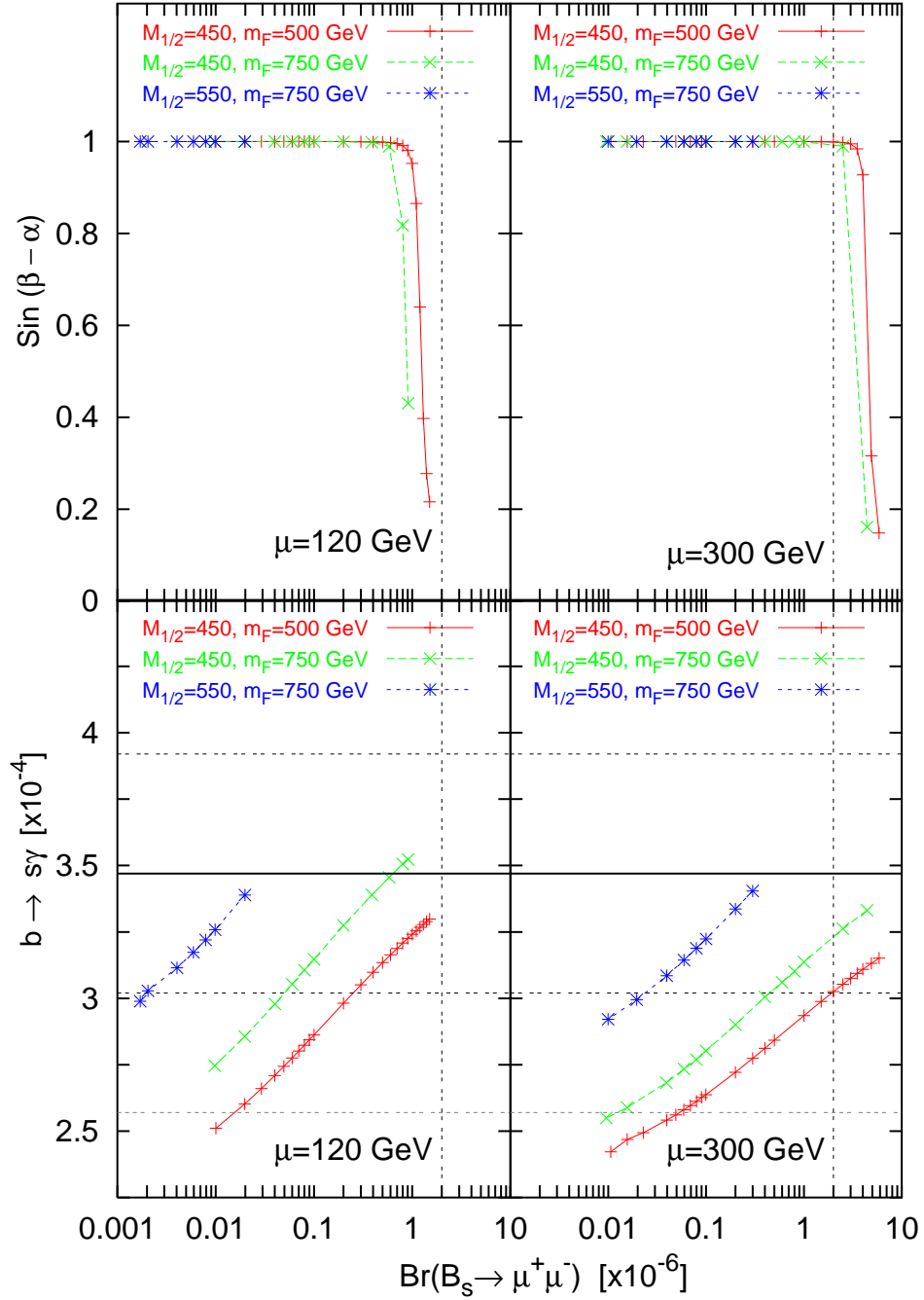}}
\begin{minipage}[t]{15cm}
\caption{\small{This figure shows the variation of
$\sin (\beta - \alpha )$ and the branching ratio for $b\to s \gamma$, 
against $B_{s}\to \mu \mu$. The vertical dashed line represents the 
present CDF
bound on $B_{s}\to\mu^{+}\mu^{-}$. The horizontal lines show the central
measured value(solid) of Br$(b\rightarrow s \gamma)$ along with the  
$1\sigma$(dashed) and $2\sigma$(light dashed) regions.}}\label{bsg_var}
\end{minipage}
\end{center}
\end{figure}

We now turn to the implications of a possible measurement 
(or an improved experimental limit) of the
branching fraction of $B_{s}\to \mu \mu$ for $SO(10)$-like models.
Fig.~\ref{bsmm_var} and \ref{bsg_var} show the effect on various
quantities of varying the branching ratio for $B_{s}\to \mu\mu$ for 
three fixed points in the $m_{F}-M_{1/2}$ plane.
 
The upper panels of fig.~\ref{bsmm_var} show the variation of $\chi^2$ as
Br$(B_{s}\to \mu\mu)$ is varied. 
As Br$(B_{s}\to \mu\mu)$ decreases the $\chi^2$ increases 
initially slowly and later rapidly. The initial slow increase
is understood from \cite{Blazek:2003wz} where it was observed that
the value of $\chi^2$ for the best fit points are insensitive to 
changes of a few GeV in the Higgs spectrum, which implies an 
insensitivity to small
changes in the branching ratio for $B_{s}\to \mu \mu$. 
Hence the points which presently
exceed the CDF bound can be forced to satisfy it with only a 
small$(\sim 0.5)$ increase in $\chi^2$. But if the bound was to be lowered
to $10^{-7}$ then this would no longer be possible with $\Delta \chi^2\sim 3$.
Hence the low $M_{1/2}$ region of the $\mu=300$ GeV plane will be ruled out
and the best fit region would move towards larger $M_{1/2}$.

The lower panels of
fig.~\ref{bsmm_var} display the variation of $m_{A^0}$ as 
Br$(B_{s}\to \mu\mu)$ is varied. As expected $m_{A^0}$ increases 
smoothly as Br$(B_{s}\to \mu\mu)$ decreases. Note the 
strong correlation of the CP-odd Higgs mass with 
Br$(B_{s}\to \mu\mu)$, which for a fixed value of $\mu$ is
quite insensitive to $m_F$ and $M_{1/2}$.

The main contribution to 
the  increase in $\chi^2$ seen in fig.~{\ref{bsmm_var}} is due to 
$b\to s \gamma$ not being fit well. The lower panels of fig.~{\ref{bsg_var}}
show the variation of Br$(b\to s \gamma)$ against 
Br$(B_{s}\to \mu\mu)$ and show a clear correlation. 
This correlation was to be 
expected as the SUSY contribution to each of these processes involves
the 23 mixings in the squark mass matrix. These panels also show why
$b \to s \gamma$ is the main contribution to the change in $\chi^2$
as the fit to $b \to s \gamma$ changes from within 1$\sigma$ to almost 
2$\sigma$. 

The upper panels of fig.~{\ref{bsg_var}} show the variation of 
$\sin (\beta - \alpha )$ as Br$(B_{s}\to \mu\mu)$ is varied. 
In the low $M_{1/2}$ region, where Br$(B_{s}\to \mu\mu)$ is near the 
current limit,
$\sin (\beta - \alpha )$ is small and hence the Z-boson
couples predominantly to the heavier CP-even Higgs $H^{0}$,
rather than the lighter Higgs $h^{0}$. 
However $\sin (\beta - \alpha )$ very quickly approaches unity as the 
Br$(B_{s}\to \mu\mu)$ decreases, corresponding to 
the standard model limit where the $h^{0}$ couples
like the standard model Higgs boson.

\section{Conclusions}

We have examined the potentially very promising signal 
$B_s\to \mu^{+}\mu^{-}$ in supersymmetry 
with large $\tan\beta\sim 50$ in a {\em top-down} approach starting
from the best fits of an $SO(10)$-like model studied recently.
Our results go beyond minimal
flavour violation investigated in previous works.
Our results show that the absolute best fits provide for the 
$B_s\to \mu^{+}\mu^{-}$ signal at the borderline of the
present limits and hence the ongoing search at the Tevatron will 
start having an impact on the global analysis of this class of 
SUSY models. 

We have discussed the implications of a measurement 
(or an improved limit) of $B_s\to \mu^{+}\mu^{-}$ 
for restricting the parameter space
of gauginos and sfermion masses, and of signals in other channels
$B_{d,s}\to \ell^{+}\ell^{-}$. We have also discussed correlations
of $B_s\to \mu^{+}\mu^{-}$ with $b\rightarrow s \gamma$ and
the CP-odd Higgs mass. An improved limit for Br$(B_{s}\to \mu\mu)$
of around $10^{-7}$ would be very restricting 
and could probe Higgs masses into the range, $m_{A^{0}}=150-300$~GeV,
with the Higgs coupling strength $\sin (\beta - \alpha )$
varying very quickly around this region.
The possible non-observation of $B_{s}\rightarrow \mu^{+}\mu^{-}$ at the 
levels suggested by our study would by no means rule out $SO(10)$-like 
models. 
In the context of the analysis in \cite{Blazek:2003wz}
this would simply highlight Minimum B, with its heavier Higgs spectrum
and Br$(B_{s}\to \mu\mu)\sim 10^{-10}$, 
as the favoured solution.
On the other hand we have seen that 
an actual observation of $B_s\to \mu^{+}\mu^{-}$ at the 
$10^{-7}$ level is quite plausibly expected 
in SUSY $SO(10)$-like models,
with interesting phenomenological and theoretical consequences.



\newpage
\clearpage
\newpage
\clearpage
\newpage
\clearpage
\newpage
\clearpage
\newpage
\clearpage

\end{document}